\documentclass[twocolumn]{article}
\usepackage{graphicx}
\usepackage{authblk}
\usepackage[left=2cm,right=2cm]{geometry}
\usepackage{amsmath, amscd, amssymb, amsthm}

\usepackage{caption}
\usepackage{subcaption}
\usepackage[title]{appendix}
\usepackage{multirow}

\title{Plasma Confinement State Classification via FPP Relevant Microwave Diagnostics}
\author[1]{Randall Clark} 
\author[2]{Vacslav Glukhov} 
\author[2]{Georgy Subbotin}
\author[2]{Maxim Nurgaliev} 
\author[2]{Aleksandr Kachkin} 
\author[3]{Max Austin} 
\author[1]{Dmitri M. Orlov}
\affil[1]{Center for Energy Research, University of California San Diego, La Jolla, CA, 92093}
\affil[2]{Next Step Fusion S.a.r.l., Luxembourg}
\affil[3]{The University of Texas at Austin, Austin, TX, 78712}

\date{August 2025}

\begin{document}


\twocolumn[
  \begin{@twocolumnfalse}
    \maketitle
    \begin{abstract}
    We present a parsimonious and robust machine learning approach for identifying plasma confinement states in fusion power plants (FPPs) where reliable identification of the low-confinement (L-mode) and high-confinement (H-mode) regimes is critical for safe and efficient operation. Unlike research-oriented devices, FPPs must operate with a severely constrained set of diagnostics. To address this challenge, we demonstrate that a minimalist model, using only electron cyclotron emission (ECE) signals, can deliver accurate and reliable state classification. ECE provides electron temperature profiles without the engineering or survivability issues of in-vessel probes, making it a primary candidate for FPP-relevant diagnostics. Our framework employs ECE as input, extracts features with radial basis functions, and applies a gradient boosting classifier, achieving high accuracy with test accuracy averaging 96\% correct predictions. Robustness analysis and feature importance study confirm the reliability of the approach. These results demonstrate that state-of-the-art performance is attainable from a restricted diagnostic set, paving the way for minimalist yet resilient plasma control architectures for FPPs.
    \end{abstract}
  \end{@twocolumnfalse}
]

\section{Introduction}
As fusion research advances toward reactor-scale devices such as FPPs, the need for new tools that operate exclusively on diagnostics expected to be viable in a reactor environment becomes increasingly urgent. Research tokamaks are equipped with a variety of diagnostic components focused on the study and control of plasmas: DIII-D alone has more than 50 systems \cite{boivin2005diii}. While these instruments are vital to both controlling the plasma and performing state estimation tasks, only a subset of them will be viable in the harsh environment of an FPP. The environment inside a future fusion reactor will impose severe constraints for any diagnostic front-end component; neutron bombardment, large heat and particle fluxes, strong electromagnetic forces, and gamma radiation all erode the sensors of the reactor over time. Additionally, physical access to these instruments will likely be limited, making it challenging to service damaged devices. 

Among the key physics parameters for stable, efficient, and safe reactor operation is the confinement regime—specifically, whether the plasma is in the low-confinement mode (L-mode) or in the high-confinement mode (H-mode). H-mode is expected to be the primary operational regime for FPPs because it introduces an edge transport barrier, forming a pedestal that increases density and temperature while reducing energy losses \cite{wagner2007quarter}. For this reason, H-mode operation underpins the physics basis of all next-generation tokamaks, including ITER, SPARC, and DEMO \cite{rodriguez2022overview,mukhovatov2003overview,siccinio2022development}. Detecting this transition accurately in real time is crucial for control and performance optimization.

Machine learning (ML) has been applied extensively in fusion research, with examples ranging from divertor detachment prediction \cite{zhu2022data,chen2025regulation,victor2024identifying}, plasma shape control \cite{degrave2022magnetic}, safety factor reconstruction \cite{wei2023reconstruction}, to H-mode detection \cite{marceca2020detection,gill2024real,matos2021plasma}. However, most existing H-mode classifiers rely on a broad set of diagnostics that will not be available in a reactor environment. This creates a critical gap for future devices.

We address this gap by developing an H-mode detection model that operates solely on diagnostics expected to remain viable in an FPP. Specifically, we demonstrate that electron cyclotron emission (ECE) measurements of the electron temperature profile — combined with minimal auxiliary signals — are sufficient to classify confinement regimes accurately and reliably. We rely on high-quality, expert-labeled data to ensure the quality of the model training.

\section{Diagnostics}
The diagnostics considered here provide the input signals for the model, with emphasis on their viability in the harsh environment of FPPs. In our approach, each diagnostic must not only supply information relevant to confinement state classification but also meet stringent engineering constraints: survivability under neutron bombardment, limited physical access, and long-term reliability with minimal maintenance.

\subsection{Electron Cyclotron Emission}


Electron cyclotron emission (ECE) provides the electron temperature profile, the most informative input for our classifier. In a magnetized plasma, radiation at electron cyclotron harmonics lies in the microwave range ($\sim$100 GHz for most tokamaks). At low harmonics, plasmas are optically thick and emit approximately as a blackbody, with power proportional to the local electron temperature $T_e$ and only weakly dependent on other plasma parameters. Because the emission frequency depends on the magnetic field strength ($B\propto 1/R$ in a tokamak), each detected frequency maps to a specific radial location, enabling reconstruction of $T_e(R,t)$.


ECE has several engineering advantages for FPP deployment. Like other microwave diagnostics, radiation is collected at the plasma boundary and transmitted via low-loss waveguides to detectors located outside high-radiation zones. No components are exposed to the plasma interior, and required port sizes are modest (50–200 mm). At reactor-relevant temperatures ($T_e > 20$ keV), relativistic effects broaden ECE harmonics, but these can be corrected with established wave modeling techniques \cite{bartlett1996}. These attributes make ECE one of the most robust and feasible diagnostics for FPPs.

ECE is already a standard technique in research tokamaks. For example, DIII-D employs a 40-channel heterodyne radiometer measuring second-harmonic ECE across the midplane \cite{austin2003,hartfuss2013mwdiag}. This provides reliable core-to-edge coverage, except at very high densities where propagation cutoffs can obscure the signal. Such conditions can be identified using independent density measurements, and we explicitly include these operational limits in our modeling framework.

\subsection{CO\textsubscript{2} Interferometer}

A CO\textsubscript{2} interferometer supplies line-integrated electron density, serving both as a basic input and as a safeguard for ECE validity. The method relies on the phase shift of a laser beam passing through the plasma; the shift is proportional to the line-integrated density \cite{van2004phase}. The system operates entirely outside the vacuum vessel, with minimal plasma intrusion, and is therefore compatible with reactor environments; however, there does remain an active question regarding if the plasma facing mirror can survive. ITER will have a CO2 interferometer called TIP to test this very issue \cite{van2017tests,van2018tests,van2013conceptual}. In our framework, the interferometer primarily informs whether ECE measurements remain within their operational envelope.

\subsection{Magnetic Diagnostics}
\subsubsection{The Toroidal Magnetic Field (BT) Measurement}
The calculation of the ECE chord locations is dependent upon the strength of the magnetic field, which is dominated by the toroidal magnetic field. We make a simplifying assumption that the toroidal component of the magnetic field is sufficient to determine the ECE chord locations for use in a data-driven model. The toroidal field is calculated purely from the toroidal field coil currents, using the equation $B_{\phi}=\mu_0I_{TF}/2\pi R$, which in turn are safely measured with Rogowski loops outside the reactor. \cite{strait2006magnetic}

\subsubsection{The Plasma Boundary Measurement}
The model also relies on an accurate determination of the plasma boundary, specifically the radial location of the outermost flux surface. The outer midplane radius of the last closed flux surface, $R_{\mathrm{mid}}^{\mathrm{out}}$, represents the last location where an ECE chord can provide a meaningful contribution to the electron temperature profile. At DIII-D $R_{\mathrm{mid}}^{\mathrm{out}}$ is estimated using EFIT \cite{lao1985reconstruction}, which reconstructs the plasma boundary based on magnetic diagnostics \cite{strait2006magnetic}.

For reactor applications, however, the long-term reliability of magnetic sensors is uncertain \cite{biel2019diagnostics,testa2010magnetic,entler2019prospects}. Optical methods for boundary reconstruction have been explored as alternatives; however, their viability is limited by radiation-induced degradation \cite{hommen2014real, shu2016plasma, todd2014diagnostic}. Ultimately, whichever diagnostic approach proves most robust will be necessary for both plasma shape control and for providing the boundary information required by our model.
\section{Label Generation}
\subsection{Hand Label Generation Process}
For the development of the model, 300 shots were selected across the 2025 and 2024 campaigns at DIII-D. Shots with the line integrated density measured by the CO\textsubscript{2} interferometer exceeding the ECE cutoff value are excluded. We also exclude shots that are part of repeated experiments on the same day to reduce correlation within the dataset and improve the balance of the state space coverage.

The expert labeling process comprised multiple steps. The primary tools for labeling were the CO\textsubscript{2} interferometer line-integrated density measurement and the Thomson Scattering (TS) electron temperature and density profiles. Sharp changes in the line-integrated density indicated L-H and H-L transitions, which were verified by the presence of pedestals in the TS profiles. Supplemental metrics, such as visible filter scopes, NBI power, ECH power, and changes in plasma parameters, including stored energy, $\beta_N$, internal inductance, and energy confinement time, were also referenced. The labels were then validated with the notes from the DIII-D electronic logbook, where experts record their observations about specific plasma events during a plasma discharge.

\subsection{Labeled Data Analysis}
The expert-provided labels were recorded as time frames labeled as either L- or H-mode, providing the user a continuous time region to choose time slices of data at their discretion. The training and test data sets were constructed, starting at 100 ms and incrementing by 100 ms, until either the data ran out or the plasma current dropped below 200 kA (indicating the end of the discharge).

Curation of the data has been performed to remove data points that would inhibit the ECE from properly measuring the electron temperature profile. The first curation is done on data that has exceeded the density cutoff limit; CO\textsubscript{2} line-integrated density measurements greater than $4.5 \cdot 10^{13}$~cm$^{-3}$ are removed. The second curation is done on data points where the ECE chords don't observe the pedestal. The ECE chords aren't required to exceed past the outer plasma boundary, $R_{\mathrm{mid}}^{\mathrm{out}}$, but must be within some distance $R_{\Delta edge}$; the hyper parameter, $R_{\Delta edge}$, was found to work best at 0.1 m where model fidelity was the least inhibited and as much data could be retained as possible. 

Curation of the data reduced the data set from 300 shots to 283. The final data set had 7668 remaining data points with 37\% H-mode labels and 63\% L-mode labels.

\begin{figure}[!ht]
\centering
\includegraphics[width=8.5cm,height=8.5cm,keepaspectratio]{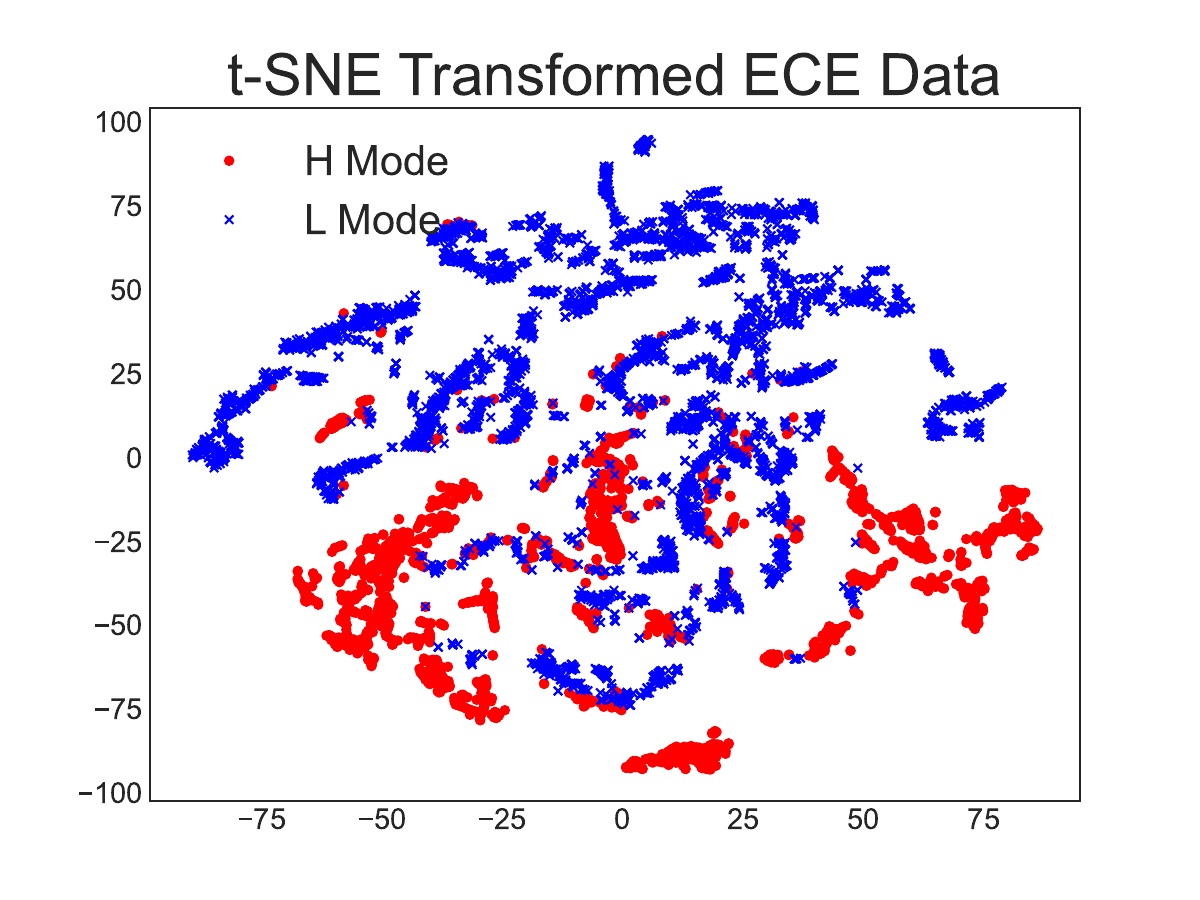}
\caption{This figure shows the breakdown of L-mode and H-mode data points in the data unprocessed 40 chord data set. The t-SNE algorithm was utilized to visualize the ECE data space and how it naturally separates out the L mode and H mode.}
\label{LH_TSNSE}
\end{figure}

\begin{figure}[!ht]
\centering
\includegraphics[width=8.5cm,height=8.5cm,keepaspectratio]{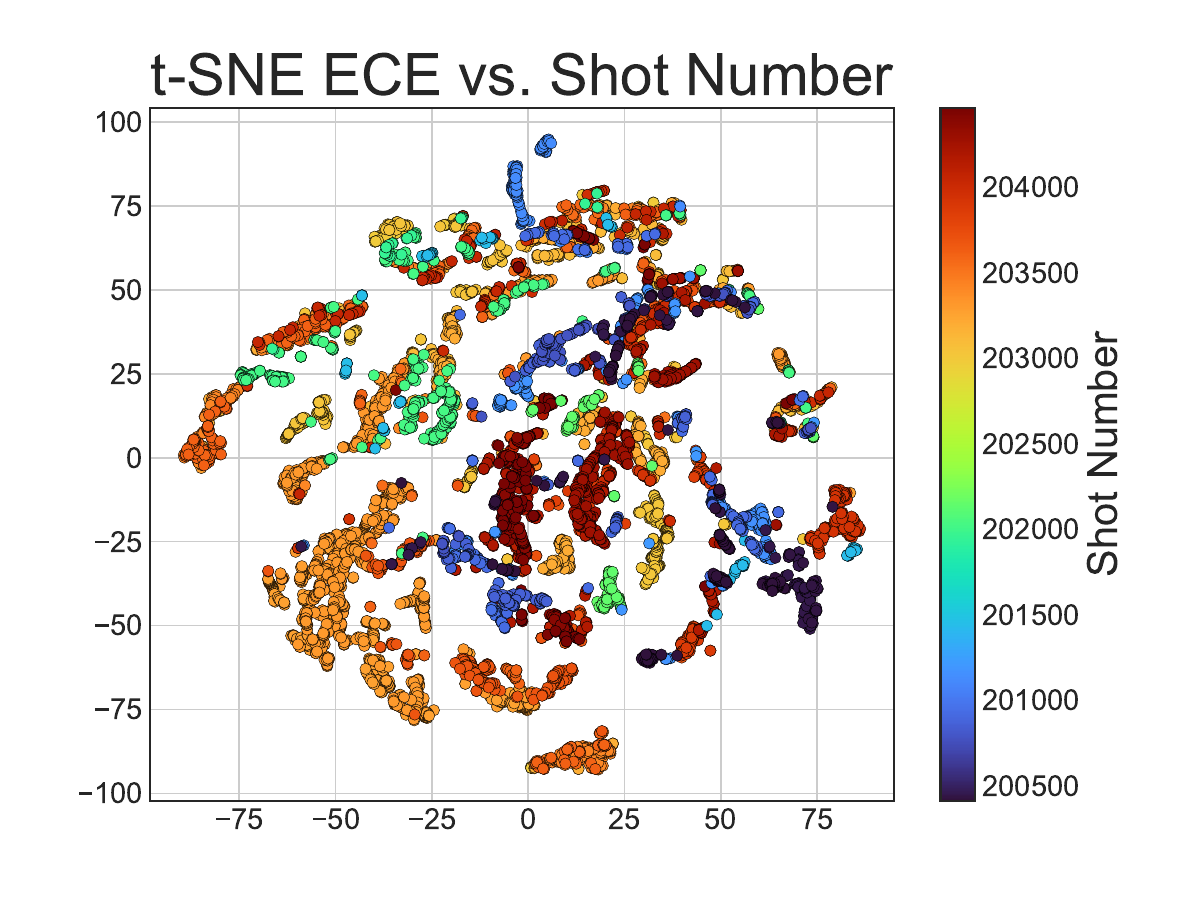}
\caption{This figure shows the same t-SNE plot as in figure \ref{LH_TSNSE}, but with a color bar indicating the different shot number values. This plot illustrates that the ECE data space is well sampled by the fact that repetitive ECE behavior is seen by overlapping points of data that occurred weeks or months apart during different physics experiments.}
\label{ShotNumber_TSNSE}
\end{figure}

\begin{figure*}[ht!]
\centering
\includegraphics[width=17cm,height=17cm,keepaspectratio]{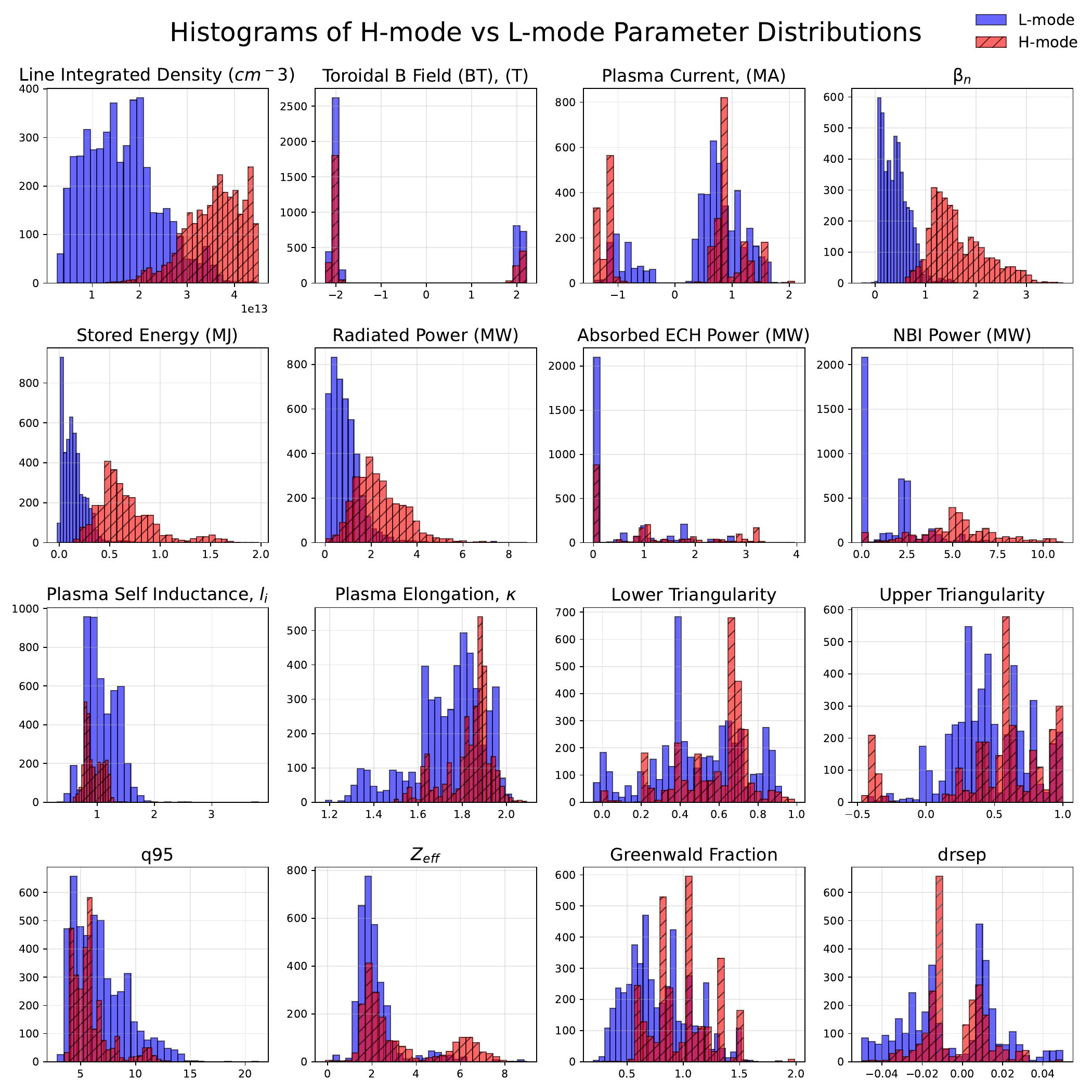}
\caption{A histogram of plasma parameters, power inputs, and shape parameters from the labeled dataset are presented here to quantify the data space. Each data point was taken from the shot at the time of training/testing (if available). The term, drsep, refers to the outboard radial distance to the second separatrix; negative values of drsep describe Lower Single Null (LSN), positive values describe Upper Single Null (USN), and values near zero describe a Double Null configurations.}
\label{Histogram}
\end{figure*}

To determine whether the data from the ECE signal is sufficient to distinguish between plasma confinement modes, we utilize the heuristic t-SNE visualization tool, which converts the full-width 40-chord signal into a 2-dimensional chart, roughly preserving the mutual relationships between data points. Figure \ref{LH_TSNSE} shows that the ECE signal likely capable of separating L-mode and H-mode. Figure \ref{ShotNumber_TSNSE} shows that the shots from different time periods exhibit a diverse set of behaviors; the mixing of shot numbers that are far apart can be indicative that the ECE data space is well sampled enough to catch some repeating patterns. A histogram of various plasma confinement mode related parameters is provided in Figure \ref{Histogram} to quantify how well sampled the dataset is; the figure can also help identify regions where the ML model may fail should the model be applied to plasmas that lie outside the sampled data space.

\section{The Foundation of the Classifier}

Stability, robustness, and safety requirements of FPPs are unlikely to be provided by traditional data-driven black box models. They lack interpretability and transparency, and cannot provide performance guarantees when applied to new data. In contrast, our data-driven model is grounded in the H-mode physics. The key difference between the two modes is the sharp gradient of temperature at the edge of the plasma, the pedestal. We use an economical representation of the temperature profile with radial basis functions (RBFs) which interpolate the profile. We then apply a binary classifier to the resulting RBF weights.

\subsection{The Radial Basis Function for Profile Fitting}
Radial Basis Functions are powerful interpolation tools that are easy to fit with the ordinary least squares method \cite{lowe1988multivariable}. The form the RBF interpolation takes is a sum of RBFs, $\psi$, with $N_c$ centers, $\boldsymbol{c}(q)$, corresponding to different locations in the data:
\begin{equation}
f(\boldsymbol{x}) = \sum_{q=1}^{N_c} \omega_q \phi(||\boldsymbol{x}-\boldsymbol{c}(q)||)
\end{equation}
The weights, $\omega_q$, are fitted to the data and the centers, $\boldsymbol{c}(q)$, are evenly spaced radially from the location of the first allowed ECE chord (near the core) to the plasma boundary, $R_{\mathrm{mid}}^{\mathrm{out}}$, in order to capture the full electron temperature profile. The RBF itself can be any variety of function, but the gaussian form is standard and generally robust so it is used (for a detailed exploration of RBFs, see reference \cite{wu2012using}):
\begin{equation}
\phi(||\boldsymbol{x}-\boldsymbol{c}(q)||) = e^{-k(||\boldsymbol{x}-\boldsymbol{c}(q)||)^2}
\end{equation}
The hyperparameter $k$, classically $1/2\sigma^2$, is set to $10^2$.

To ensure the electron temperature profile is represented correctly, the following preprocessing steps are applied to the ECE data:
\begin{itemize}
    \item Remove all chords outside the plasma, beyond $R_{\mathrm{mid}}^{\mathrm{out}}$ as they aren't measuring plasma.
    \item Assume the chord with the highest temperature corresponds to the plasma core.
    \item Discard inner chords (chords with a smaller radius than the assumed plasma core) only after the temperature falls below a percentage of the maximum, $T_{\%max}$ (typically 90\% of the maximum temperature).
    \item Add a synthetic data point at $R_{\mathrm{mid}}^{\mathrm{out}}$ with a temperature value of zero, as the temperature should be negligible outside the plasma boundary.
\end{itemize}

\begin{figure}[!ht]
\centering
\includegraphics[width=8.5cm,height=8.5cm,keepaspectratio]{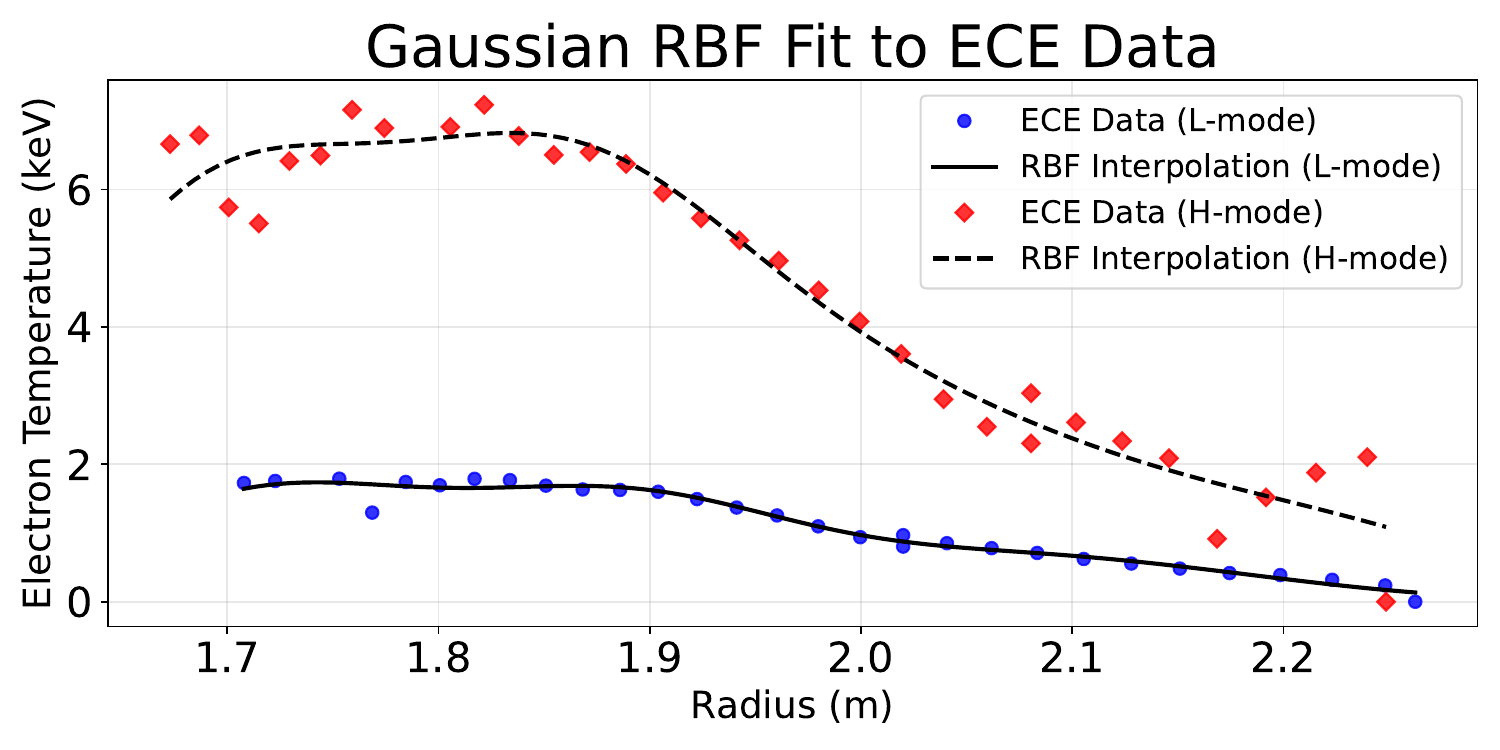}
\caption{ECE data from L-mode shot 203047 at time 4500 ms and H-mode shot 203191 at time 2400 ms are each interpolated with 7 Gaussian RBFs and plotted. The 7 centers from the RBFs were placed equally distant apart, one at each end and 5 equally spaced within.}
\label{RBF_Fit_ECE_Plot}
\end{figure}

Figure \ref{RBF_Fit_ECE_Plot} shows an example of ECE data during an H-mode shot being interpolated by seven RBFs with chords ranging from the plasma core to the outer edge of the plasma. The purpose for the interpolation is to embed the information of the electron temperature profile into the RBF weights. Each weight contains information regarding both the temperature and location along the 1D profile with the most significant information stored in the weights near the edge region where pedestals emerge in H-mode.

\subsection{The Binary Classifier}
\begin{figure}[!ht]
\centering
\includegraphics[width=8.5cm,height=8.5cm,keepaspectratio]{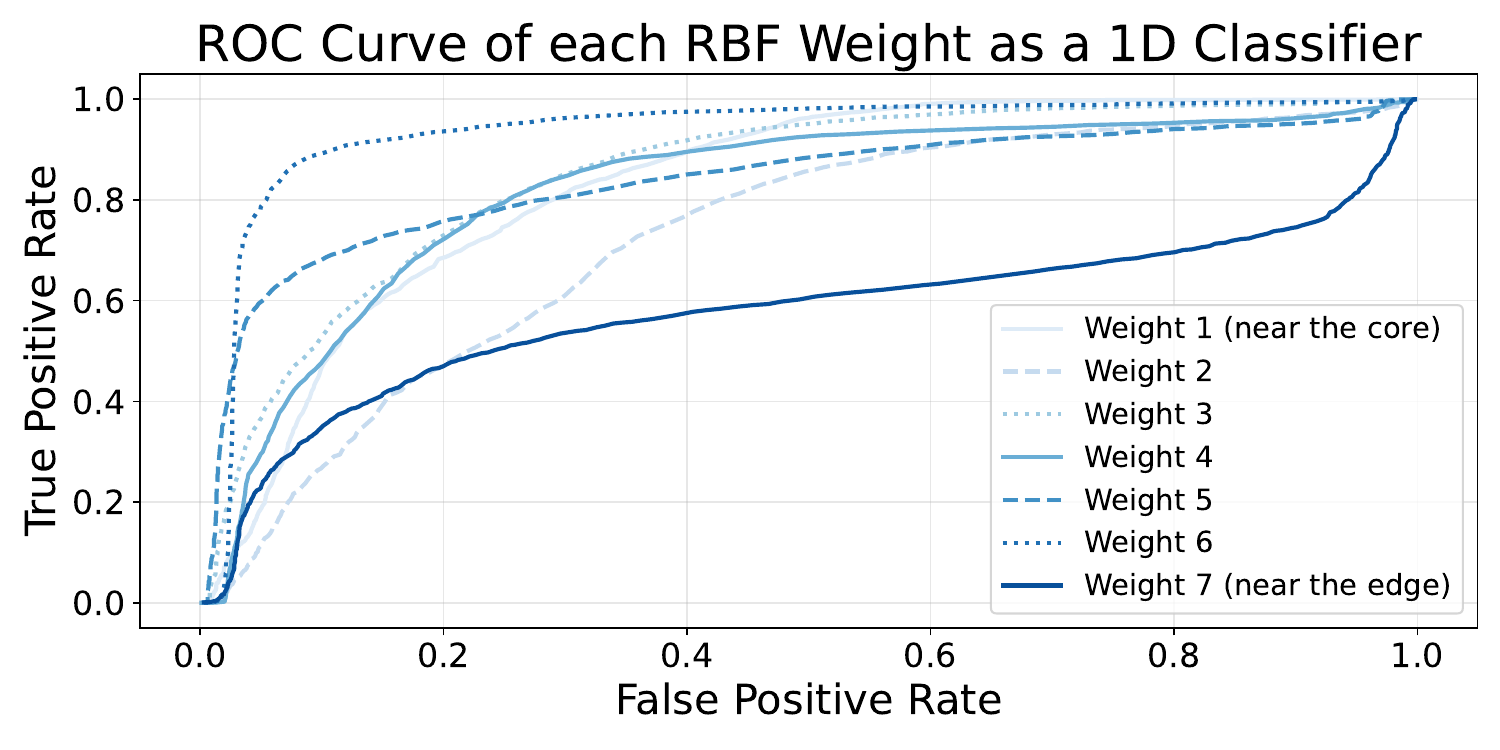}
\caption{An ROC Curve of each of the 7 center weight pairings is made using a simple decision stump classifier that was carefully scaled across the data set. These curves illustrate which regions of the profile provide the most discriminative information for mode classification.}
\label{ROC_Curve}
\end{figure}

With the physics information embedded in the RBF weights, they can be used to make a physics informed binary classifier. A simple binary classifier would be to take each of the seven weights individually and to make a basic 1D decision stump classifier from them. This exploratory step aims to assess which RBF weights provide the most discriminative information, rather than to achieve high accuracy. With this stump classifier, an ROC curve is plotted in figure \ref{ROC_Curve}. What this figure tells us is that the most significant center for separating the L-modes from H-modes is the center one over from the plasma edge, the one that most directly observes the pedestal.

\begin{figure}[!ht]
\centering
\includegraphics[width=8.5cm,height=8.5cm,keepaspectratio]{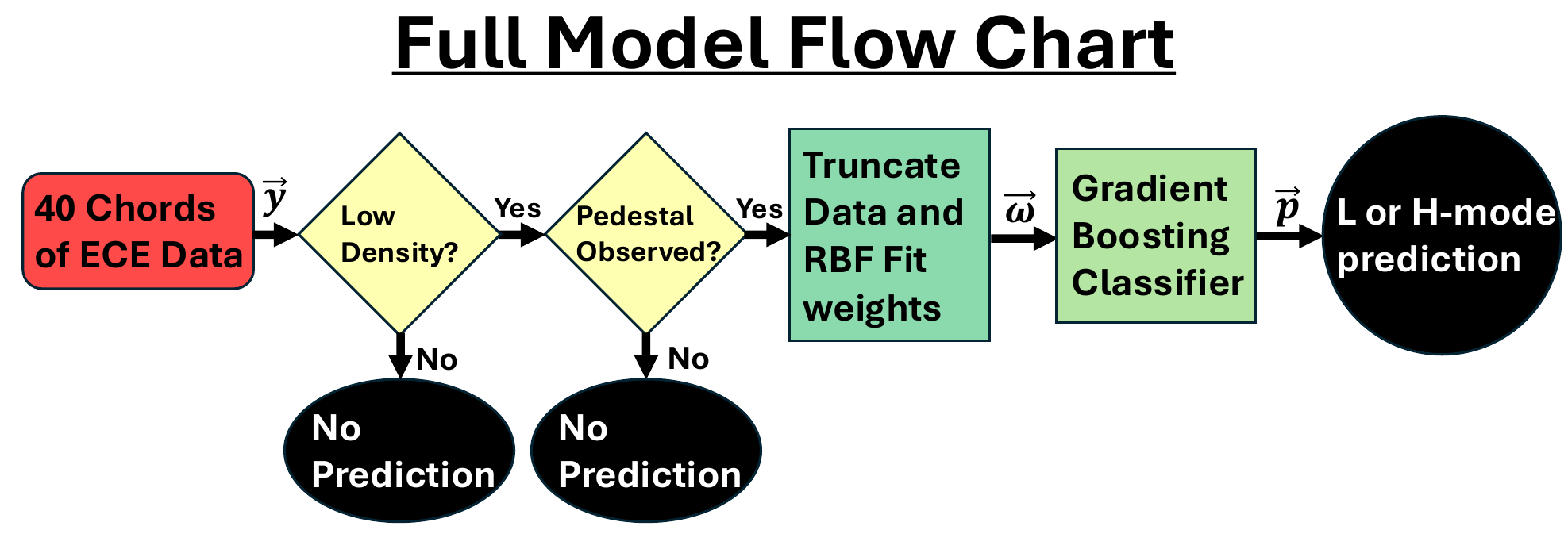}
\caption{This flow chart gives a visual representation of how the model goes from ECE data to making a prediction of the plasma confinement state.}
\label{FlowChart}
\end{figure}

Experiments with several different binary classifiers were executed including decision trees, support vector machines, and Gradient Boosting Classifiers (GBC) and found that gradient boosting, specifically the HistGradientBoostingClassifier from sklearn, to be the most effective for this data set and quick to train \cite{scikitlearn,friedman2001greedy}. The full set of hyper parameters used for data filtering, RBF fitting, and training the GBC is listed out on Table \ref{GBC_HyperParam}. A visual description of the full process is displayed in figure \ref{FlowChart}.

\begin{table}[!ht]
\begin{center}
\begin{tabular}{||c c||} 
 \hline
 Model Parameter & Value  \\ [0.5ex] 
 \hline\hline
 Start Time  & 100 ms  \\ 
 \hline
 Time Increment   & 100 ms  \\ 
 \hline
 Total Shots  & 283  \\ 
 \hline
  Train/Test Split  & 80/20  \\ 
 \hline
 Total Data Points  & 7668  \\ 
 \hline
 L:H ratio   & 4806:2862  \\ 
 \hline
  Density Limit  & $4.5 \cdot 10^{13}$~cm$^{-3}$   \\ 
  \hline
  $R_{\Delta edge}$  & 0.1 m  \\ 
  \hline
  $T_{\%max}$  & 90\%  \\ 
 \hline
 No. RBF Centers   & 7  \\ 
 \hline
 Gaussian RBF parameter   &  $10^2$ \\ 
 \hline
 Learning Rate & 0.1  \\ 
 \hline
 Max Iterations & 100 \\
 \hline
  Max Leaf Nodes & 31  \\ 
 \hline
  L2 Regularization & 0  \\ 
 \hline
\end{tabular}
\end{center}
\caption{Table of Parameters used for building and curating the data set, fitting the RBFs, and training the GBC.}
\label{GBC_HyperParam}
\end{table}

\section{Model Results}
After extensive testing and tuning of hyperparameters, the model performance achieves an average test accuracy of  96\% with a 95\% F1 score across 100 reshuffles of test and train shots. A full description of the model performance can be found in table \ref{GBCModelPerformance}. This accuracy is comparable to other H-mode classifiers in other works\cite{gill2024real,matos2021plasma,orozco2022neural}.

\begin{table}[!ht]
\begin{center}
\begin{tabular}{||c c c||} 
 \hline
 GBC Model & Average & Standard Deviation \\ [0.5ex] 
 \hline\hline
 Test Accuracy  & 96\% & 0.90\% \\ 
 \hline
 Test Precision  & 94\% & 2.3\% \\ 
 \hline
 Test Recall  & 96\% & 1.8\% \\ 
 \hline
 Test F1  & 95\% & 1.4\% \\ 
 \hline
\end{tabular}
\end{center}
\caption{The statistics of the GBC model using RBF weights to predict H-mode are calculated from 100 reshuffles of the shot numbers in the test and train data set. The test/train split was performed by shot rather than by time steps to prevent temporal leakage, which would otherwise inflate test accuracy due to highly correlated data within individual shots.}
\label{GBCModelPerformance}
\end{table}

\begin{figure}[!ht]
\centering
\includegraphics[width=8.5cm,height=8.5cm,keepaspectratio]{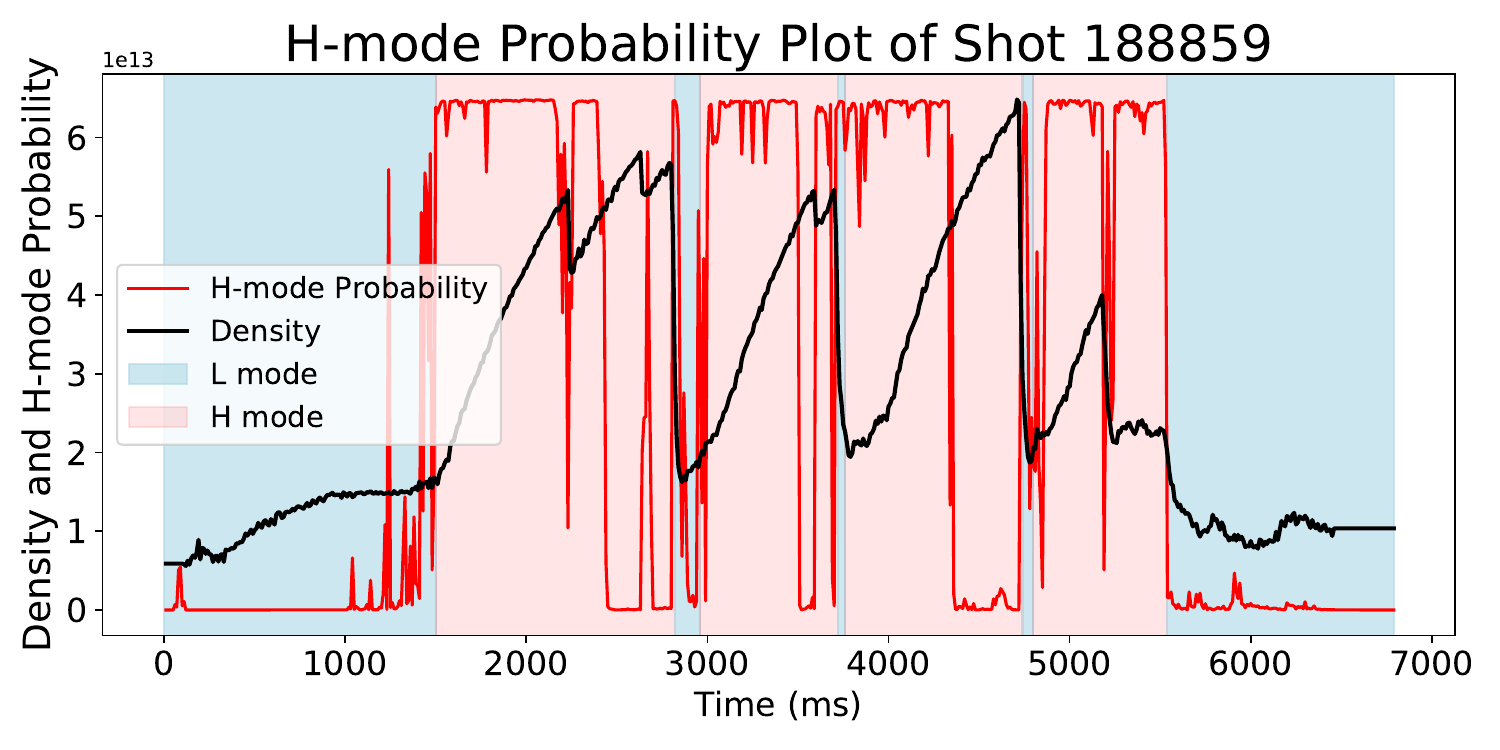}
\caption{Model predictions for shot 188859 show high confidence in H-mode classification except during brief intervals when the density exceeded $5.0 \cdot 10^{13}$~cm$^{-3}$ degrading the ECE signal quality. Shot 188859 is an USN plasma that was run in early 2024, months before the first data point in the training data set.}
\label{PlotShot188859}
\end{figure}

\begin{figure}[!ht]
\centering
\includegraphics[width=8.5cm,height=8.5cm,keepaspectratio]{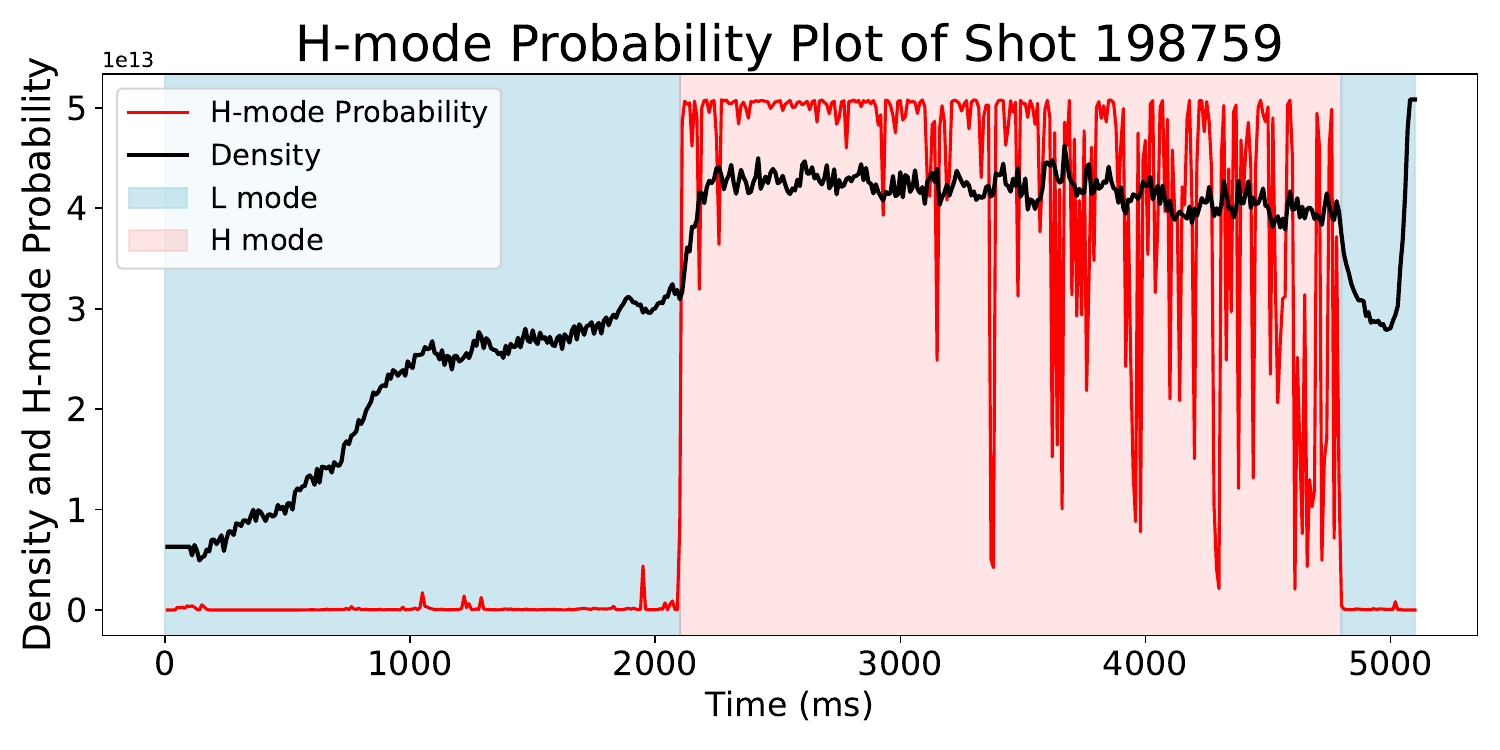}
\caption{The model recognized when the plasma entered and exited H-mode. The model confidence during the latter half of H-mode began to deteriorate  due to the strong oscillation observed. This was a LSN plasma.}
\label{PlotShot198759}
\end{figure}

\begin{figure}[!ht]
\centering
\includegraphics[width=8.5cm,height=8.5cm,keepaspectratio]{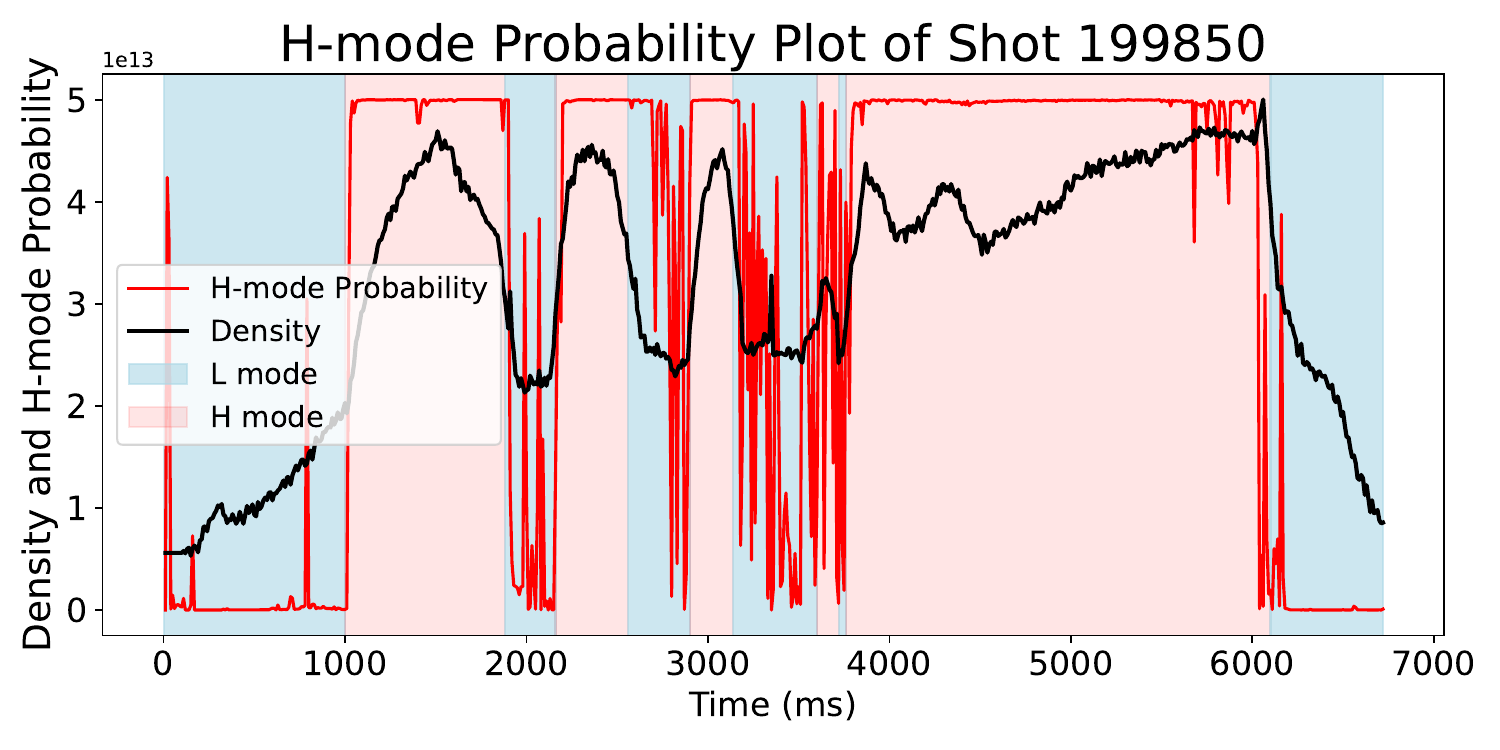}
\caption{Shot 199850 exhibited multiple L-H and H-L transitions, all of which were correctly detected. Model confidence decreased after each back-transition to L-mode, likely due to the plasma exhibiting behavior suggestive of an imminent return to H-mode. Shot 199850 was a LSN plasma.}
\label{PlotShot199850}
\end{figure}

\begin{figure}[!ht]
\centering
\includegraphics[width=8.5cm,height=8.5cm,keepaspectratio]{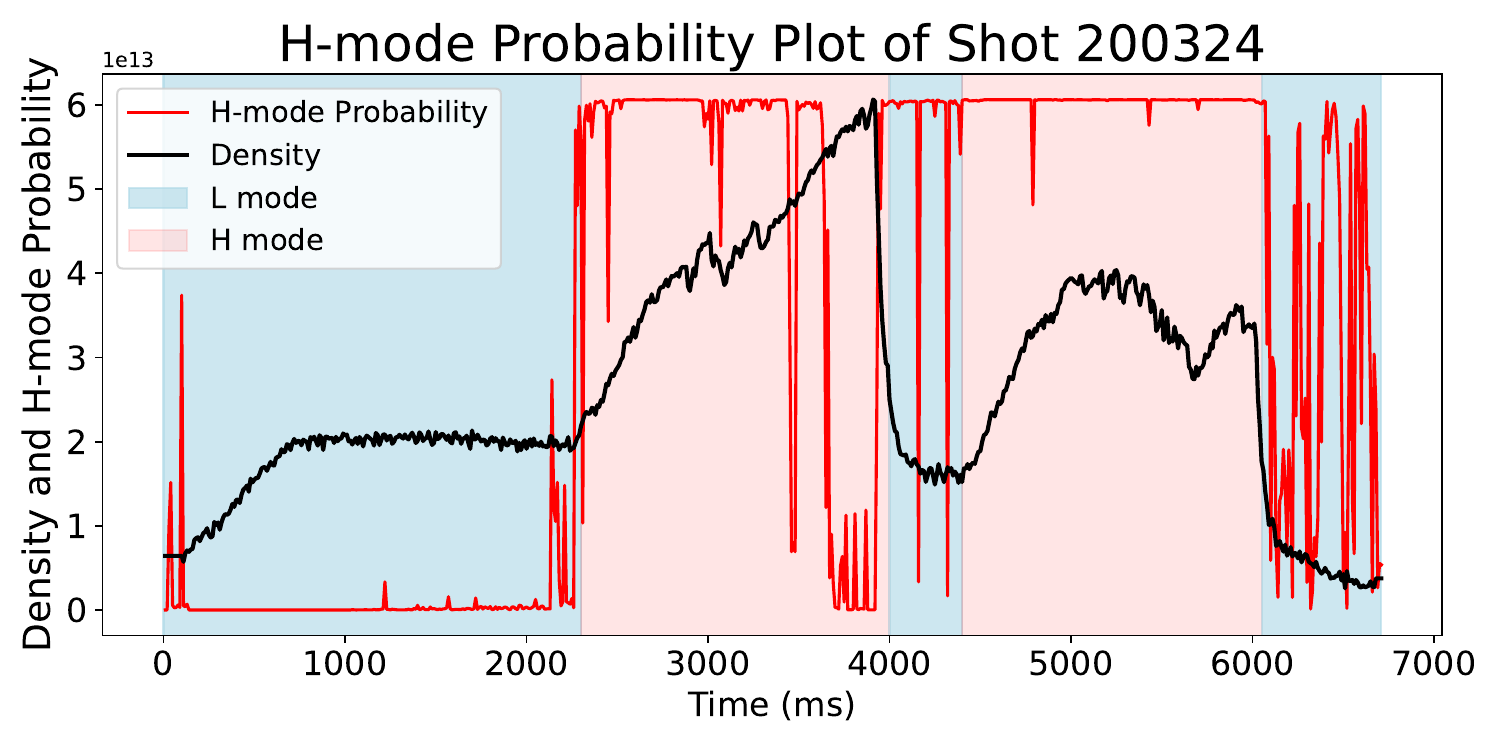}
\caption{Shot 200324 provides another example that demonstrates the effect of the density limit on model accuracy. Shot 200324 was a LSN plasma.}
\label{PlotShot200324}
\end{figure}

To evaluate the ability of the model to accurately classify shots not found in either the training or test set, the model was applied to four shots from outside the original labeled dataset. These four shots all have ECE chords that observe the pedestal region throughout the entire shot and minimally exceed the density limit. The results in figures \ref{PlotShot188859}, \ref{PlotShot198759}, \ref{PlotShot199850}, and \ref{PlotShot200324} present the model's predicted probability of H-mode instead of a binary classification to provide insight into the model’s confidence over time. It can be seen in these figures that unusual behavior, like the oscillations in figure \ref{PlotShot198759}, can lower model confidence. The figures also illustrate the impact of the density limit, as predictions degrade once the line-averaged density exceeds $5.0 \cdot 10^{13}$~cm$^{-3}$.

\subsection{Model Robustness} 
Robustness — the ability of a system to function, or at least fail gracefully, under adverse or unexpected conditions —is a core requirement for all high-stakes, safety-critical domains. Fusion plasma diagnostic and control systems in FPPs are no exception. In contrast, data-driven models not explicitly designed for robustness often prove fragile, particularly under data loss or contamination. Our model addresses this challenge by incorporating radial basis function feature extraction, which is expected to remain resilient to partial data loss and noisy inputs. In this section, we test that assumption.

Diagnostic degradation and occasional failure are inevitable over an instrument’s lifetime. While ECE is generally reliable, individual channels can still fail. In research environments, the worst-case rate is about one failed channel per 1,000 plasma shots, with faults typically corrected before another occurs; simultaneous multi-channel loss is therefore rare. In FPPs, however, continuous operation and limited maintenance access amplify both the likelihood and the cost of failures. Probing the model’s sensitivity to chord dropout is therefore not only a robustness test but also a guide to diagnostic design: it quantifies the classifier’s tolerance margins. It also identifies the minimum number of channels required to ensure reliable performance in reactor conditions.

\begin{figure}[!ht]
\centering
\includegraphics[width=8.5cm,height=8.5cm,keepaspectratio]{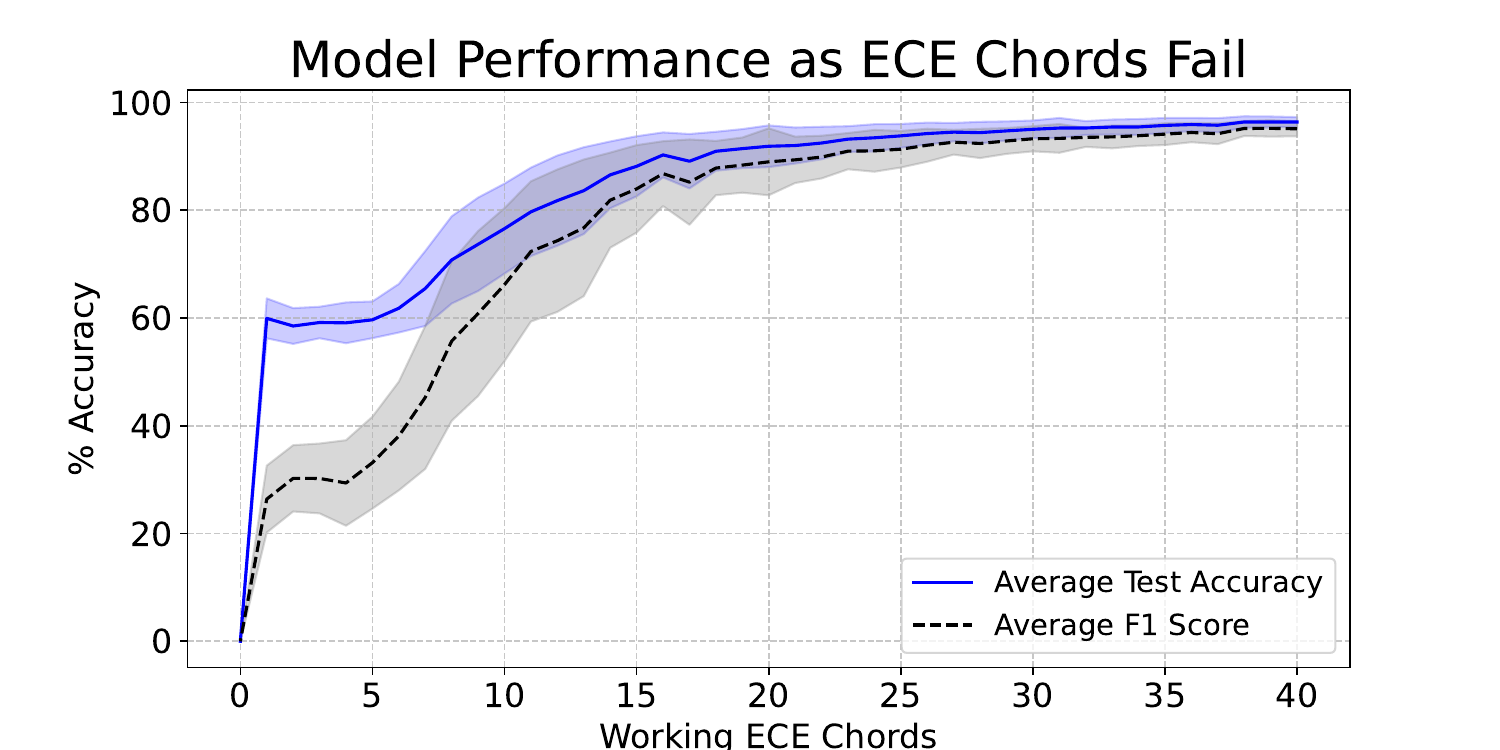}
\caption{To test the robustness of the GBC model on RBF weights to loss of ECE chords, ECE chords are removed from the test data set at random at a particular failure rate. An ensemble of 100 randomization tests is performed at each failure rate, with the model trained on all 40 chords. The same chords are excluded across all test data for each ensemble iteration before reshuffling.}
\label{ChordFailureRandom}
\end{figure}

Figure \ref{ChordFailureRandom} shows the average classification accuracy as a function of the number of active ECE chords. The model maintains high accuracy with as few as 15 channels; below this threshold, performance deteriorates sharply. The large standard deviations highlight that not all chords are equally informative. Figure \ref{ROC_Curve} further demonstrates that channels sampling the pedestal region contribute most strongly to confinement state classification.

In practice, noise rather than complete chord failure may be the more relevant challenge. For a heterodyne radiometer, the minimum noise level is set by the photon noise, which depends on the local electron temperature as
\[
\Delta T_e = T_e \sqrt{\frac{2 BW_v}{BW_{if}}},
\]
where $BW_v$ and $BW_{if}$ are instrument bandwidths. Because of this scaling, core channels with higher $T_e$ values are intrinsically noisier than edge channels. In realistic conditions, however, electronic photon noise can be kept at the $\sim 1\%$ level by using narrow bandwidth amplifiers, and the dominant fluctuations instead arise from plasma phenomena themselves (e.g., ELMs near the edge, MHD modes at mid-radius, or sawtooth oscillations in the core).

\begin{figure}[!ht]
\centering
\includegraphics[width=8.5cm,height=8.5cm,keepaspectratio]{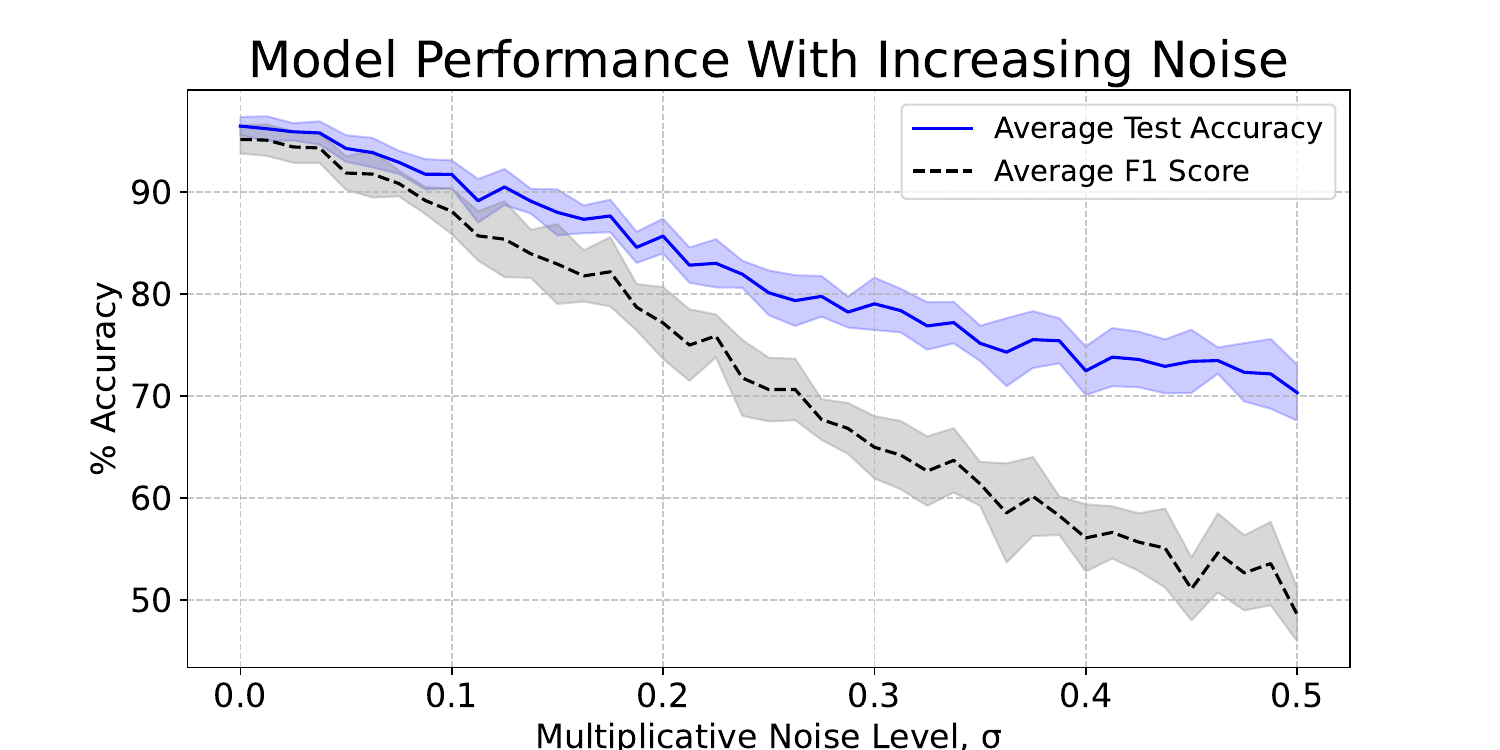}
\caption{Using a model trained on clean data, it is applied to the noisy test data over an ensemble of iterations. Using the multiplicative noise model, Eq \ref{MultitplicativeNoiseModel}, the variance $\sigma$ is scaled up, showing a decrease in test accuracy. The error bars represent the standard deviation.}
\label{NoiseModel}
\end{figure}

Using a multiplicative noise model, the test data can simulate the temperature scaled noise as photon noise would. Equation \ref{MultitplicativeNoiseModel} shows how noise is added to the data through multiplication of the measured temperature and the log-normal distribution:
\begin{equation}\label{MultitplicativeNoiseModel}
\hat{T} = T*\exp{(\mathcal{N}(-\sigma^2/2,\sigma^2))}
\end{equation}
All test data would be scaled by this noise model, where $\mathcal{N}$ is the normal distribution, and $\sigma$ is the variance of the normal distribution. Figure \ref{NoiseModel} depicts the model performance as the variance/noise of the model increases. The mean is chosen as $-\sigma^2/2$ so that that the expectation value of the lognormal distribution is one.

\section{Conclusion and future work}
We have developed a parsimonious and robust data-driven tool for H-mode classification based solely on electron cyclotron emission (ECE) signals. The model achieves an average test accuracy of 96\% and an F1 score of 95\%, and remains reliable even under significant input signal loss. Its minimalist design facilitates integration into fusion power plant environments and ensures adaptability to evolving operating conditions.

Looking ahead, additional diagnostics relevant to FPPs—such as reflectometry for electron density profiles and interferometer–polarimetry for density fluctuations—should be evaluated not only for predictive value but also for integration and maintenance costs. Incorporating such diagnostics within a hybrid identification framework would allow ancillary models to act as fallbacks in the event of primary sensor failure or diagnostic limitations, thereby strengthening the overall robustness of plasma state control systems.
\section{Acknowledgement}
This material is based upon work supported by the U.S. Department of Energy, Office of Science, Office of Fusion Energy Sciences, using the DIII-D National Fusion Facility, a DOE Office of Science user facility, under Awards DE-FC02-04ER54698, DE-FG02-05ER54809, DE-FG02-97ER54415, and Next Step Fusion S.a.r.l. with UCSD staff supported by Next Step Fusion S.a.r.l.  The authors would like to thank Wilkie Choi and Michael Van Zeeland for fruitful discussions.
\section*{Disclaimer}
This report was prepared as an account of work sponsored by an agency of the United States Government. Neither the United States Government nor any agency thereof, nor any of their employees, makes any warranty, express or implied, or assumes any legal liability or responsibility for the accuracy, completeness, or usefulness of any information, apparatus, product, or process disclosed, or represents that its use would not infringe privately owned rights. Reference herein to any specific commercial product, process, or service by trade name, trademark, manufacturer, or otherwise does not necessarily constitute or imply its endorsement, recommendation, or favoring by the United States Government or any agency thereof. The views and opinions of authors expressed herein do not necessarily state or reflect those of the United States Government or any agency thereof.

\bibliographystyle{plain}

\end{document}